\documentclass[letterpaper,american,english,aip]{revtex4-1}
\usepackage[T1]{fontenc}
\usepackage{textcomp}
\usepackage[utf8]{inputenc}
\usepackage[active]{srcltx}
\usepackage{amsmath}
\usepackage{amssymb}

\makeatletter

\usepackage{babel}
\usepackage{xcolor}
\colorlet{BLUE}{blue}

\newcommand{\new}[1]{{\color{red}#1}}
\let\FVoldcite\cite
\renewcommand{\cite}[1]{{\color{black}\FVoldcite{#1}}}
\let\FVoldref\ref
\renewcommand{\ref}[1]{{\color{black}\FVoldref{#1}}}
\let\FVoldpageref\pageref
\renewcommand{\pageref}[1]{{\color{black}\FVoldpageref{#1}}}
\let\FVoldeqref\eqref
\renewcommand{\eqref}[1]{{\color{black}\FVoldeqref{#1}}}

\makeatother

\begin{document}
\title{Feshbach-Villars Formalism for a Spin-1/2 Particle in Curved Spacetime}

\author{Abdelmalek Boumali}
\email{abdelmalek.boumali@univ-tebessa.dz or boumali.abdelmalek@gmail.com}

\affiliation{Echahid Cheikh Larbi Tebessi University, Tebessa, Algeria}
\date{\today}
\begin{abstract}
This study extends the Feshbach--Villars (FV) formalism to encompass spin--$1/2$ particles propagating in curved spacetime. Starting from the covariant Dirac equation, we derive its Hamiltonian form using a tetrad (vierbein) formulation and implement a diagonal FV transformation on the associated second-order (squared) equation. The resulting Schr\"odinger-type evolution equation is expressed in both block-matrix and Pauli-matrix representations, explicitly displaying gravitational and electromagnetic couplings, alongside spin--field interaction terms. We analyze this formalism in $(1+2)$- and $(1+3)$-dimensional settings and apply it to cosmic-string backgrounds (static and spinning, with and without oscillator couplings) to determine the corresponding energy spectra and exact wave functions. Although the FV reformulation is algebraically equivalent to the covariant Dirac dynamics---subject to the standard restriction that eliminates spurious solutions introduced by squaring---it offers a highly practical complementary perspective. Specifically: (i) the particle and antiparticle sectors are distinctly represented by two coupled components equipped with a simple (pseudo-)unitary inner product tied to the conserved charge; (ii) the physical distinction between stationary and static spacetimes is directly encoded in the FV operator $\mathrm{Y}$, which is governed by the ADM shift vector, thus clearly diagnosing how frame-dragging modifies particle--antiparticle mixing; and (iii) the FV block structure provides a streamlined approach for solving spectral problems in topologically nontrivial geometries. \new{For the spinning string with a screw dislocation, the analysis is carried out at the metric (Levi--Civita) level, the defect entering through the curved tetrad; a dedicated appendix establishes that an explicit Riemann--Cartan treatment retaining the contortion term yields the same effective angular shift and exterior radial operator, and that, under the same regular-at-origin quantization, it reproduces the same closed-form spectrum.}
\end{abstract}

\maketitle

\section{Introduction}

A Hamiltonian (first-order-in-time) formulation is frequently the most practical starting point for relativistic quantum dynamics, particularly when investigating conserved inner products, spectral problems, and the separation of particle and antiparticle sectors. For spin-$0$ particles, the Feshbach--Villars (FV) construction offers a canonical method to recast the second-order Klein--Gordon equation into a Schr\"odinger-type evolution system featuring an indefinite (charge-type) norm \cite{Feshbach1958}. This perspective has proven conceptually and computationally advantageous, seeing extensive use in both analytical and numerical studies \cite{Merad2000,Bounames2001,Haouat2005,Guettoul2006,Brown2015,Motamedi2019,Wingard2024a}.

For spin-$1/2$ particles, a comparable FV-type linearization can be derived from the \emph{squared} Dirac operator (the Schr\"odinger or second-order form of the Dirac equation). In flat spacetime, this FV--$1/2$ construction and its pseudo-unitary structure were thoroughly developed and discussed by Robson and Staudte \cite{Robson1996,Staudte1996}; see also the thesis by Garcia Vallejo \cite{GarciaVallejoThesis} and more recent pedagogical treatments \cite{Brown2015,Wingard2024a}. Because this material serves as a foundational standard, Section II briefly summarizes it to establish notation and render the subsequent curved-spacetime generalization transparent.

To extend the FV framework to general relativistic settings, Silenko \cite{Silenko2008,Silenko2013} identified a conformal invariance of the massless scalar theory in Riemannian spacetimes---namely, an invariance of the properly defined wave equation under local Weyl rescalings $g_{\mu\nu}\to\Omega^{2}(x)g_{\mu\nu}$ (accompanied by a corresponding field rescaling). This property restricts curvature couplings and supports a Hamiltonian formulation with well-behaved transformation properties. Silenko derived Hamiltonian forms for both massive and massless scalar particles in arbitrary spacetimes and established their classical limits. Recently, this methodology has been successfully applied across various contexts \cite{Bouzenada2023a,Garah2025EPJC,Garah2025NPB,Bouzenada2023b,Bouzenada2023c,Bouzenada2024}. Conversely, FV-type Hamiltonian formulations for \emph{spinors} in curved spacetime remain less developed. This is partly because the relevant second-order operator incorporates both curvature and spin-coupling terms, and partly because stationary (non-static) geometries introduce mixing structures that are absent in purely diagonal metrics.

The primary objective of this work is to establish a systematic FV--$1/2$ Hamiltonian formulation for spin-$1/2$ particles in curved spacetime, rooted directly in the standard covariant Dirac equation. The intent is not to postulate a novel relativistic wave equation, but rather to construct a convenient Hamiltonian representation within an expanded FV state space where: (a) the separation into positive- and negative-frequency sectors is explicit; (b) the conserved inner product is naturally articulated using a straightforward metric operator; and (c) the influence of geometric data---particularly the lapse and shift (ADM) decomposition---can be explicitly tracked at the operator level. These advantages become especially evident in stationary defect geometries featuring nontrivial $g_{0\varphi}$ or $g_{z\varphi}$ components, where the FV operator $\mathrm{Y}$ is non-vanishing and the azimuthal quantum number experiences shifts due to energy- and momentum-dependent terms (see Section V).

The primary contributions of this paper are as follows:
\begin{itemize}
\item A curved-spacetime FV--$1/2$ Hamiltonian derived from the covariant Dirac equation in a tetrad (vierbein) framework, featuring explicit gravitational and electromagnetic couplings alongside spin--field interaction terms.
\item An operator-level demonstration of how \emph{stationarity} (a non-zero ADM shift) is incorporated into the FV formalism via the FV operator $\mathrm{Y}$, thereby diagnosing particle--antiparticle mixing in frame-dragging backgrounds.
\item Applications to cosmic-string geometries (both static and spinning, with and without oscillator couplings), yielding explicit energy spectra and wave functions in both $(1+2)$ and $(1+3)$ dimensions. This includes the characteristic energy dependence induced by $g_{0\varphi}\neq 0$.
\end{itemize}

\new{We emphasize that the novelty of this work lies not in the energy spectra themselves---which, on the physical subspace, necessarily coincide with those of the covariant Dirac equation---but in the Hamiltonian reformulation itself. Three aspects are central: (i) the construction of a Feshbach--Villars spin-$1/2$ Hamiltonian directly in curved spacetime, in which the particle and antiparticle sectors and the conserved (pseudo-)unitary inner product are made explicit; (ii) the careful identification of the physical Dirac subspace that must be retained after squaring, which guarantees equivalence with the first-order theory and removes the spurious solutions; and (iii) the operator-level role of the FV operator $\mathrm{Y}$, governed by the ADM shift vector, as a diagnostic that distinguishes stationary from static geometries and tracks frame-dragging in the particle--antiparticle decomposition. These features are not transparent in the standard first-order Dirac formulation, and it is in this structural and diagnostic sense that the present formalism is complementary to it.}

It is important to stress the scope of the present work in light of the well-known fact that squaring the Dirac operator introduces additional solutions. Throughout this paper, the FV evolution is interpreted as a Hamiltonian representation of Dirac dynamics restricted to the physical (Dirac) subspace; this restriction eliminates spurious solutions and ensures equivalence with the original first-order Dirac equation (see Section IV).

\paragraph*{Comparison with the Standard Dirac Equation}
The FV--$1/2$ formalism used here is not a new relativistic wave equation independent of the Dirac theory. Rather, it is a Hamiltonian reformulation of the \emph{squared} Dirac equation within an enlarged state space. The standard Dirac equation is first order in both time and space derivatives and acts on a four-component spinor in $(1+3)$ dimensions. In contrast, the FV--$1/2$ representation is first order in time but second order in space, acting on an eight-component object $\Psi_{FV}=(\phi,\chi)^T$, where each entry is itself a Dirac spinor. Consequently, the two descriptions are algebraically equivalent only after restricting the FV dynamics to the physical Dirac subspace, thereby excising the spurious solutions introduced by the squaring process. The primary advantage of the FV formulation is therefore not the generation of new spectra on the physical subspace, but rather a structural reorganization: particle and antiparticle sectors are explicitly delineated, the conserved bilinear form is encoded by a simple metric operator, and the role of stationary off-diagonal metric terms emerges transparently through the operator $\mathrm{Y}$ \cite{Feshbach1958,Bounames2001,Silenko2013,Wingard2024a}.

The remainder of this paper is organized as follows. Section II reviews the FV linearization for the second-order Dirac equation in flat spacetime and establishes the notation. Section III derives the Hamiltonian form of the covariant Dirac equation in curved spacetime, with an emphasis on tetrad fields and the spin connection. Section IV develops the FV formulation for spin--$1/2$ particles in general curved backgrounds and discusses the transformation properties of the resulting Hamiltonian. Section V applies the formalism to cosmic-string geometries (including oscillator couplings) and derives the corresponding energy spectra. Finally, Section VI provides a dedicated discussion of our findings alongside concluding remarks outlining the scope, limitations, and potential extensions of the proposed approach.

\section{Review of the Feshbach-Villars Formalism for Spin-1/2 Particles in Flat Spacetime}

\subsection{Theoretical Framework}

This section compiles standard background material regarding the second-order Dirac equation and its FV-type Hamiltonian reformulation in flat spacetime. This presentation primarily serves to establish conventions and notation; for detailed derivations and extensive discussions, we refer the reader to Refs.~\cite{Robson1996,Staudte1996,GarciaVallejoThesis,Brown2015,Wingard2024a}.

We begin by recalling the standard covariant Dirac equation governing a fermion of mass $m$ coupled to an external electromagnetic potential \cite{Greiner2001}:
\begin{equation}
\left(i\gamma^{\mu}D_{\mu}-m\right)\psi=0
\label{1}
\end{equation}
where $D_{\mu}=\partial_{\mu}+\frac{iq}{\hbar c}A_{\mu}$ represents the gauge-covariant derivative, and $A_{\mu}$ is the electromagnetic four-potential. To construct a second-order wave equation, the conjugate operator is applied to both sides:
\begin{equation}
\left(i\gamma^{\nu}D_{\nu}+m\right)\left(i\gamma^{\mu}D_{\mu}-m\right)\psi=0
\label{2}
\end{equation}
Expanding this operator product yields:
\begin{equation}
\left[(i\gamma^{\nu}D_{\nu})(i\gamma^{\mu}D_{\mu})-m^{2}\right]\psi=0
\label{3}
\end{equation}
which can be equivalently expressed as:
\begin{equation}
-\gamma^{\nu}\gamma^{\mu}D_{\nu}D_{\mu}\psi-m^{2}\psi=0
\label{4}
\end{equation}
By exploiting the fundamental Clifford algebra satisfied by the Dirac matrices:
\begin{equation}
\{\gamma^{\mu},\gamma^{\nu}\}=2g^{\mu\nu}
\label{5}
\end{equation}
and introducing the antisymmetric spin tensor, the differential operator can be separated into symmetric and antisymmetric contributions:
\begin{equation}
\gamma^{\nu}\gamma^{\mu}=g^{\nu\mu}+\sigma^{\nu\mu}
\label{6}
\end{equation}
where
\begin{equation}
\sigma^{\nu\mu}=\frac{1}{2}[\gamma^{\nu},\gamma^{\mu}]
\label{7}
\end{equation}
Consequently, the second-order derivative term decomposes as follows:
\begin{equation}
-\gamma^{\nu}\gamma^{\mu}D_{\nu}D_{\mu}=-g^{\nu\mu}D_{\nu}D_{\mu}-\sigma^{\nu\mu}D_{\nu}D_{\mu}
\label{8}
\end{equation}
The first term corresponds to the gauge-covariant d'Alembertian, whereas the second term introduces the coupling to the electromagnetic field tensor $F_{\mu\nu}$ through the commutator:
\begin{equation}
[D_{\nu},D_{\mu}]=iqF_{\nu\mu},\quad F_{\nu\mu}=\partial_{\nu}A_{\mu}-\partial_{\mu}A_{\nu}
\label{9}
\end{equation}
This transforms the spin-interaction term into:
\begin{equation}
-\sigma^{\nu\mu}D_{\nu}D_{\mu}=\frac{iq}{2}\sigma^{\nu\mu}F_{\nu\mu}
\label{10}
\end{equation}
Consolidating these results yields the generalized second-order Dirac equation:
\begin{equation}
\left[D^{\mu}D_{\mu}-m^{2}+\frac{iq}{2}\sigma^{\nu\mu}F_{\nu\mu}\right]\psi=0
\label{11}
\end{equation}
Expanding this electromagnetic coupling in terms of the electric ($\mathbf{E}$) and magnetic ($\mathbf{B}$) fields yields the well-known Feynman--Gell-Mann equation \cite{Brown2015,Wingard2024a}:
\begin{equation}
\left[D^{\mu}D_{\mu}-m^{2}+\frac{q}{\hbar c}\left(\mathbf{B}\cdot\boldsymbol{\sigma}-i\mathbf{E}\cdot\boldsymbol{\sigma}\right)\right]\psi=0.
\label{12}
\end{equation}

\subsection{Feshbach-Villars Linearization}

To establish a Hamiltonian time-evolution formalism, the FV procedure partitions the spinor state into two auxiliary components, $\phi$ and $\chi$:
\begin{equation}
\psi=\phi+\chi
\label{13}
\end{equation}
\begin{equation}
\left(i\frac{\partial}{\partial t}-V\right)\psi=m(\phi-\chi)
\label{14}
\end{equation}
where $V=qA_0$ represents the scalar potential. Substituting this ansatz transforms the second-order wave equation into a pair of coupled first-order equations:
\begin{equation}
\left(i\frac{\partial}{\partial t}-V\right)(\phi+\chi)=m(\phi-\chi)
\label{15}
\end{equation}
\begin{equation}
\left(i\frac{\partial}{\partial t}-V\right)(\phi-\chi)=\frac{p^{2}}{m}(\phi+\chi)+m(\phi+\chi)-\frac{i}{m}(\nabla V\cdot\boldsymbol{\sigma})(\phi+\chi)
\label{16}
\end{equation}
This pair of relations can be consolidated into a two-component block-matrix Schr\"odinger-like equation:
\begin{equation}
i\frac{\partial}{\partial t}\begin{pmatrix}\phi\\
\chi
\end{pmatrix}=H\begin{pmatrix}\phi\\
\chi
\end{pmatrix}
\label{17}
\end{equation}
The effective FV Hamiltonian is given by \cite{Wingard2024a,Brown2015}:
\begin{equation}
H=\begin{pmatrix}1 & 1\\
-1 & -1
\end{pmatrix}\frac{p^{2}}{2m}+\begin{pmatrix}1 & 0\\
0 & -1
\end{pmatrix}m+\begin{pmatrix}1 & 0\\
0 & 1
\end{pmatrix}V-\begin{pmatrix}1 & 1\\
-1 & -1
\end{pmatrix}\frac{i}{2m}(\nabla V\cdot\boldsymbol{\sigma})
\label{18}
\end{equation}
Alternatively, expressing this Hamiltonian structure using the Pauli matrices $\tau_{i}$ acting on the FV component space yields:
\begin{equation}
H=(\tau_{3}+i\tau_{2})\frac{p^{2}}{2m}+\tau_{3}m+I_{2}V-(\tau_{3}+i\tau_{2})\frac{i}{2m}(\nabla V\cdot\boldsymbol{\sigma})
\label{19}
\end{equation}
where $I_{2}$ represents the $2\times2$ identity matrix.

\subsection{The FV Oscillator for Spin-1/2 Particles}

To investigate the Dirac oscillator within the FV framework, we introduce a non-minimal coupling by applying the substitution $p \to p - im\omega x$ to the momentum operator. For simplicity in this introductory flat-space example, we restrict our analysis to a one-dimensional spatial coordinate $x$. In this context, $\omega$ represents the characteristic frequency of the oscillator.

Implementing this generalized momentum, the corresponding one-dimensional free FV Hamiltonian (assuming a vanishing scalar potential, $V=0$) takes the form \cite{Bouzenada2023a}:
\begin{equation}
H=(\tau_{3}+i\tau_{2})\frac{\left(p+im\omega x\right)\left(p-im\omega x\right)}{2m}+\tau_{3}m
\label{20}
\end{equation}
Consequently, substituting this structure into the coupled FV system yields the Schr\"odinger-like formulation governing the one-dimensional Dirac oscillator:
\begin{equation}
\begin{aligned}
\frac{1}{2m}\left[\frac{d^{2}}{dx^{2}}-m^{2}\omega^{2}x^{2}+m\omega\right]\left(\phi+\chi\right)+m\phi & =E\phi\\
-\frac{1}{2m}\left[\frac{d^{2}}{dx^{2}}-m^{2}\omega^{2}x^{2}+m\omega\right]\left(\phi+\chi\right)-m\chi & =E\chi
\end{aligned}
\label{21}
\end{equation}
Applying standard algebraic reductions analogous to those used for the free FV system, we deduce the following decoupled second-order differential equation for the combined wave function $\psi = \phi + \chi$:
\begin{equation}
\left[\frac{d^{2}}{dx^{2}}-m^{2}\omega^{2}x^{2}+\left(E^{2}-m^{2}+m\omega\right)\right]\psi(x)=0
\label{22}
\end{equation}
This relation can be expressed more compactly as:
\begin{equation}
\left[\frac{d^{2}}{dx^{2}}-\lambda_{1}x^{2}+\lambda_{2}\right]\psi(x)=0
\label{23}
\end{equation}
where we have defined the auxiliary parameters:
\begin{equation}
\lambda_{1}=m^{2}\omega^{2},\quad\lambda_{2}=E^{2}-m^{2}+m\omega.
\label{24}
\end{equation}
Imposing standard normalizability boundary conditions yields the corresponding energy quantization condition:
\begin{equation}
\frac{E^{2}-m^{2}+m\omega}{2m}=\left(n+\frac{1}{2}\right)\omega,\quad n=0,1,2,\ldots
\label{25}
\end{equation}
Rearranging this constraint provides the explicit analytical expression for the discrete energy spectrum:
\begin{equation}
E=\pm\sqrt{m^{2}+2m\omega\left(n+\frac{1}{2}\right)}
\label{26}
\end{equation}
This outcome aligns with established results for the quantum oscillator. Such equivalence is physically anticipated for this reduced equation: the observable spectrum must concur once the projection onto the physical sector is performed.

\subsection{Distinctions Between the FV-1/2 Oscillator and the Standard Dirac Oscillator}

It is instructive to explicitly delineate what changes---and what remains invariant---when transitioning from the standard Dirac oscillator to its FV--$1/2$ representation. In the conventional Dirac approach, one begins with a first-order equation,
\begin{equation}
\left[i\gamma^{\mu}D_{\mu}-m\right]\Psi=0,
\end{equation}
where the spin degrees of freedom enter directly through the matrix structure of the Hamiltonian. In contrast, the FV--$1/2$ approach first squares the Dirac operator and subsequently linearizes the resulting second-order equation in time. This procedure yields an eight-component Schr\"odinger-type system featuring explicit particle and antiparticle blocks. Consequently:
\begin{enumerate}
\item The \emph{physical spectrum} must match the Dirac spectrum once the Dirac constraint is enforced;
\item The \emph{wave-function representation} differs, as the FV state harbors redundant components prior to the application of the physical restriction;
\item The \emph{spin coupling} manifests through the Pauli term inherited from the squared Dirac operator, rather than via the original first-order Dirac Hamiltonian.
\end{enumerate}
In the elementary one-dimensional example analyzed here, the resulting equation (Eq.~(\ref{22})) encapsulates solely the second-order oscillator structure and does not exhibit an explicit spin splitting. Consequently, the spectrum in Eq.~(\ref{26}) coincides with the Klein--Gordon-type result. This concurrence should not be construed as evidence that the spin-$1/2$ Dirac equation is identical to the spin-$0$ Klein--Gordon equation. Rather, it reflects the fact that the utilized reduction procedure temporarily suppresses the spin-resolved fine structure that naturally emerges in a full first-order Dirac oscillator analysis.

\selectlanguage{american}%

\section{Hamiltonian Form of the Dirac Equation in Curved Spacetime}

Deriving the Hamiltonian form of the Dirac equation in curved spacetime requires recasting the covariant equation into a time-evolution format governed by an operator acting on the wave function. The covariant Dirac equation in curved spacetime reads:
\begin{equation}
\left[i\gamma^{\mu}\mathcal{D}_{\mu}-m\right]\Psi=0,
\label{eq:1}
\end{equation}
where:
\begin{itemize}
\item $\gamma^{\mu}=e_{a}^{\mu}\gamma^{a}$ denote the curved-spacetime gamma matrices, with $e_{a}^{\mu}$ being the tetrad (vierbein) fields and $\gamma^{a}$ the standard flat-spacetime gamma matrices.
\item $\mathcal{D}_{\mu}=\partial_{\mu}+\frac{1}{4}\omega_{\mu}^{ab}\gamma_{ab}-iqA_{\mu}$ is the spinor covariant derivative, incorporating both the spin connection and the minimal electromagnetic coupling.
\item $m$ represents the mass of the spinor field.
\end{itemize}
The tetrad fields $e^{a}{}_{\mu}$ and their inverses $e_{a}{}^{\mu}$ relate the curved spacetime metric to the local Minkowski metric via $g_{\mu\nu}=e^{a}{}_{\mu}e^{b}{}_{\nu}\,\eta_{ab}$, thereby defining the curved gamma matrices.
It is crucial to distinguish between: (i) the standard spacetime covariant derivative $\nabla_{\mu}$ acting on tensor indices; (ii) the gauge-covariant derivative $D_{\mu}=\partial_{\mu}-iqA_{\mu}$ employed for minimal electromagnetic coupling; and (iii) the spinor covariant derivative $\mathcal{D}_{\mu}$, which includes the spin connection to guarantee local Lorentz covariance. These derivatives coincide when acting on scalar functions but differ when applied to tensors or spinors due to their respective connection coefficients \cite{Nakahara2003}.

Expanding Eq.~(\ref{eq:1}) and isolating the time derivative yields:
\begin{equation}
i\gamma^{0}\mathcal{D}_{0}\Psi=\left[-i\gamma^{i}\mathcal{D}_{i}+m\right]\Psi.
\label{eq:3}
\end{equation}
where the temporal and spatial components of the spinor derivative are separated as
\begin{equation}
\mathcal{D}_{0}=\partial_{0}+\frac{1}{4}\omega_{0}^{ab}\gamma_{ab}-iqA_0,\quad
\mathcal{D}_{i}=\partial_{i}+\frac{1}{4}\omega_{i}^{ab}\gamma_{ab}-iqA_i
\label{eq:4}
\end{equation}
and
\begin{equation}
\gamma_{ab}=\frac{1}{2}[\gamma_{a},\gamma_{b}].
\label{eq:5}
\end{equation}
Multiplying Eq.~(\ref{eq:3}) from the left by $(\gamma^{0})^{-1}$ (noting that in standard conventions $(\gamma^{0})^{-1}=\gamma^{0}$ when assuming an orthogonal time coordinate, or more generally using the inverse), we obtain:
\begin{equation}
i\partial_{0}\Psi=\gamma^{0}\left[-i\gamma^{i}\left(\partial_{i}+\frac{1}{4}\omega_{i}^{ab}\gamma_{ab}-iqA_i\right)+m-\frac{1}{4}\omega_{0}^{ab}\gamma_{ab}+iqA_0\right]\Psi.
\label{eq:6}
\end{equation}
This establishes the Hamiltonian form:
\begin{equation}
i\partial_{0}\Psi=\mathcal{H}\Psi,
\label{eq:7}
\end{equation}
where the Hamiltonian operator $\mathcal{H}$ is given by:
\begin{equation}
\mathcal{H}=\gamma^{0}\left[-i\gamma^{i}\left(\partial_{i}+\frac{1}{4}\omega_{i}^{ab}\gamma_{ab}-iqA_i\right)+m-\frac{1}{4}\omega_{0}^{ab}\gamma_{ab}+iqA_{0}\right].
\label{eq:9}
\end{equation}
This Hamiltonian governs the time evolution of the spinor field $\Psi$ in curved spacetime, accounting for both gravitational and gauge interactions.

The probability density associated with the Dirac equation in curved spacetime is intimately tied to the conserved current:
\begin{equation}
J^{\mu}=\bar{\psi}\gamma^{\mu}\psi,
\label{eq:10}
\end{equation}
where $\bar{\psi}=\psi^{\dagger}\gamma^{0}$ is the Dirac adjoint.
The covariant conservation of $J^{\mu}$ is guaranteed by the Dirac equation:
\begin{equation}
\nabla_{\mu}J^{\mu}=0,
\label{eq:11}
\end{equation}
where $\nabla_{\mu}$ is the metric-compatible covariant derivative. Consequently, the probability density $\rho$ corresponds to the time-component of the current:
\begin{equation}
\rho=J^{0}=\bar{\psi}\gamma^{0}\psi.
\label{eq:12}
\end{equation}
To properly normalize the wave function in curved spacetime, $\rho$ is integrated over a spacelike hypersurface $\Sigma$:
\begin{equation}
\int_{\Sigma}\rho\,\sqrt{-g}\,\text{d}^{3}x=1,
\label{eq:13}
\end{equation}
where $\sqrt{-g}$ incorporates the metric determinant to provide the invariant proper volume element.

\selectlanguage{english}%

\section{The Feshbach-Villars Formalism for Spin-1/2 Particles in Curved Spacetime}

\selectlanguage{american}%
The generalized second-order wave equation obtained by squaring the Dirac operator---originally derived by Erwin Schr\"odinger~\cite{Schrodinger1932}---is given by
\begin{equation}
\left(\frac{1}{\sqrt{-g}}\mathcal{D}_{\mu}\left(\sqrt{-g}g^{\mu\nu}\mathcal{D}_{\nu}\right)-\frac{1}{4}R+\frac{ie}{2}F_{\mu\nu}\sigma^{\mu\nu}+m^{2}\right)\Psi=0
\label{eq:14}
\end{equation}
where $R$ is the Ricci scalar, $F_{\mu\nu}$ is the electromagnetic field strength tensor, and $\mathcal{D}_{\mu}$ is the spinor covariant derivative:
\begin{equation}
\mathcal{D}_{\mu}=\partial_{\mu}+\frac{1}{4}\omega_{\mu}^{ab}\gamma_{ab}-iqA_{\mu}.
\label{eq:15}
\end{equation}
Following the methodology outlined by Silenko \cite{Silenko2013}, the general form of the covariant second-order equation can be expanded as (see Appendix A for details):
\begin{equation}
\left(\partial_{0}^{2}+\frac{1}{g^{00}\sqrt{-g}}\left\{ \partial_{i},\sqrt{-g}g^{0i}\right\} \partial_{0}+\frac{1}{g^{00}\sqrt{-g}}\partial_{i}\sqrt{-g}g^{ij}\partial_{j}+\frac{m^{2}-\lambda R}{g^{00}}\right)\psi=0.
\label{eq:16}
\end{equation}
Here, the curly brackets $\{\ldots,\ldots\}$ denote the anticommutator.
Expanding the d'Alembertian-like term $\mathcal{D}_{\mu}\left(\sqrt{-g}g^{\mu\nu}\mathcal{D}_{\nu}\right)$ explicitly over temporal and spatial indices yields:
\begin{align}
\mathcal{D}_{\mu}\left(\sqrt{-g}g^{\mu\nu}\mathcal{D}_{\nu}\right) & =
\mathcal{D}_{0}\left(\sqrt{-g}g^{00}\mathcal{D}_{0}\right)+\mathcal{D}_{0}\left(\sqrt{-g}g^{01}\mathcal{D}_{1}\right)+\mathcal{D}_{0}\left(\sqrt{-g}g^{02}\mathcal{D}_{2}\right)+\mathcal{D}_{0}\left(\sqrt{-g}g^{03}\mathcal{D}_{3}\right)\nonumber \\
 & +\mathcal{D}_{1}\left(\sqrt{-g}g^{10}\mathcal{D}_{0}\right)+\mathcal{D}_{1}\left(\sqrt{-g}g^{11}\mathcal{D}_{1}\right)+\mathcal{D}_{1}\left(\sqrt{-g}g^{12}\mathcal{D}_{2}\right)+\mathcal{D}_{1}\left(\sqrt{-g}g^{13}\mathcal{D}_{3}\right)\nonumber \\
 & +\mathcal{D}_{2}\left(\sqrt{-g}g^{20}\mathcal{D}_{0}\right)+\mathcal{D}_{2}\left(\sqrt{-g}g^{21}\mathcal{D}_{1}\right)+\mathcal{D}_{2}\left(\sqrt{-g}g^{22}\mathcal{D}_{2}\right)+\mathcal{D}_{2}\left(\sqrt{-g}g^{23}\mathcal{D}_{3}\right)\nonumber \\
 & +\mathcal{D}_{3}\left(\sqrt{-g}g^{30}\mathcal{D}_{0}\right)+\mathcal{D}_{3}\left(\sqrt{-g}g^{31}\mathcal{D}_{1}\right)+\mathcal{D}_{3}\left(\sqrt{-g}g^{32}\mathcal{D}_{2}\right)+\mathcal{D}_{3}\left(\sqrt{-g}g^{33}\mathcal{D}_{3}\right).
\label{eq:17}
\end{align}
Dividing the entire equation by $g^{00}$, we obtain the isolated time-evolution form:
\begin{equation}
\left(D_{0}^{2}+\frac{1}{g^{00}\sqrt{-g}}\left\{ D_{i},\sqrt{-g}g^{0i}\right\} D_{0}+\frac{1}{g^{00}\sqrt{-g}}D_{i}\sqrt{-g}g^{ij}D_{j}+\frac{\frac{ie}{2}F_{\mu\nu}\sigma^{\mu\nu}}{g^{00}}+\frac{m^{2}-\frac{R}{4}}{g^{00}}\right)\psi=0,
\label{eq:18}
\end{equation}
which represents a direct spinor generalization of the scalar equation.

To construct the FV Hamiltonian for this generalized second-order equation, we apply the Feshbach-Villars transformation. Recalling the expanded equation:
\begin{equation}
\left(D_{0}^{2}+\frac{1}{g^{00}\sqrt{-g}}\left\{ D_{i},\sqrt{-g}g^{0i}\right\} D_{0}+\frac{1}{g^{00}\sqrt{-g}}D_{i}\sqrt{-g}g^{ij}D_{j}+\frac{\frac{ie}{2}F_{\mu\nu}\sigma^{\mu\nu}}{g^{00}}+\frac{m^{2}-\frac{R}{4}}{g^{00}}\right)\psi=0,
\label{eq:19}
\end{equation}
where $D_{\mu}=\partial_{\mu}+\Gamma_{\mu}$ represents the relevant covariant derivative.
The FV transformation reformulates the wave function $\psi$ into two distinct components, $\phi$ and $\chi$:
\begin{equation}
\psi=\phi+\chi,\quad i\left(D_{0}+\mathrm{Y}\right)\psi=m(\phi-\chi),
\label{eq:20}
\end{equation}
where the operator $\mathrm{Y}$ is defined as:
\begin{equation}
\mathrm{Y}=\frac{1}{2g^{00}\sqrt{-g}}\left\{ D_{i},\sqrt{-g}g^{0i}\right\}.
\label{eq:21}
\end{equation}

\selectlanguage{english}%

\subsection{The Feshbach-Villars Metric and Inner Product}

The FV transformation defined in Eq.~(\ref{eq:20}) recasts the \emph{squared Dirac equation} (Eq.~(\ref{eq:14})) into a first-order Schr\"odinger-type evolution equation operating within an enlarged state space. Specifically, it introduces the two-component FV field $\Psi_{FV}=(\phi,\chi)^{T}$.
For spin-$1/2$ particles, the components $\phi$ and $\chi$ are themselves Dirac spinors (e.g., four-component objects in $(1+3)$ dimensions), meaning that $\Psi_{FV}$ comprises eight components in total.

Unlike standard Dirac theory, the FV transformation is generally not unitary with respect to the conventional positive-definite $L^{2}$ inner product.
Instead, it is pseudo-unitary with respect to an indefinite metric operator associated with the conserved \emph{charge} of the underlying second-order theory.
In the canonical FV basis, this metric operator is defined as $\eta_{FV}=\tau_{3}\otimes\mathbb{I}_{4}$:
\[
\eta_{FV}=\tau_{3}\otimes\mathbb{I}_{4}
=\begin{pmatrix}
\mathbb{I}_{4} & 0\\
0 & -\mathbb{I}_{4}
\end{pmatrix}.
\]
The corresponding conserved inner product on a hypersurface $\Sigma_{t}$ of constant coordinate time $t$ is given by:
\begin{equation}
\langle\Psi_{1}|\Psi_{2}\rangle_{FV}=\int_{\Sigma_{t}}d^{3}x\sqrt{h}\Psi_{1}^{\dagger}\eta_{FV}\Psi_{2},
\end{equation}
where $h=\det(h_{ij})$ is the determinant of the induced spatial metric on $\Sigma_{t}$.
For the FV-$1/2$ framework based on Eq.~(\ref{eq:14}), this bilinear form serves as the natural spinorial generalization of the scalar FV charge and is strictly conserved under time evolution.
Provided the initial data are restricted to valid solutions of the underlying first-order Dirac equation (Eq.~(\ref{eq:1})), the FV charge aligns consistently with the standard conserved Dirac current (Eq.~(\ref{eq:12})).
Consequently, the Hamiltonian $\mathcal{H}$ derived within this framework is not Hermitian in the standard sense ($\mathcal{H}^{\dagger}\neq \mathcal{H}$), but is instead \textit{pseudo-Hermitian} with respect to $\eta_{FV}$.
This pseudo-Hermiticity is expressed mathematically as:
\[
\mathcal{H}^{\dagger}=\eta_{FV}\mathcal{H}\eta_{FV}^{-1},
\]
which guarantees that the time-evolution operator is pseudo-unitary, thereby preserving the generalized indefinite inner product \cite{Mostafazadeh2002,Silenko2008}.

\subsection{Equivalence with the Dirac Equation and Elimination of Spurious Solutions}

Because the FV Hamiltonian originates from the \emph{squared} Dirac operator, it is imperative to address a well-known mathematical subtlety: while every solution of the first-order Dirac equation inherently satisfies the squared equation, the converse is generally false.
The second-order equation admits additional, spurious solutions that violate the original first-order Dirac constraint.
In this study, the FV formulation is anchored to Eq.~(\ref{eq:1}). We therefore treat the FV Hamiltonian evolution as a convenient Hamiltonian \emph{representation} of Dirac dynamics operating within the physical subspace.
Operationally, given an FV state $\Psi_{FV}=(\phi,\chi)^{T}$, the physical Dirac spinor is reconstructed via $\Psi=\phi+\chi$, while the auxiliary combination $\phi-\chi$ is deterministically fixed by the first-order constraint in Eq.~(\ref{eq:20}).
If the initial data satisfy the Dirac constraint, the FV evolution preserves this constraint, so that no spurious solutions are generated.
This ``Dirac restriction'' formally establishes the equivalence between the FV-$1/2$ formalism and the standard covariant Dirac theory.

\selectlanguage{american}%
\subsection{Matrix Representation of the Hamiltonian}

Using the FV transformation, the second-order equation is reduced to a system of coupled first-order equations in time.
Substituting the definitions of $\phi$ and $\chi$, we formulate the time-evolution equation:
\begin{equation}
i\partial_{0}\Psi_{FV}=\mathcal{H}\Psi_{FV},
\label{eq:22}
\end{equation}
To express the Hamiltonian operator $\mathcal{H}$ properly, let us define the effective spatial kinetic operator $\mathcal{O}$, which incorporates the generalized spatial derivatives along with the electromagnetic spin coupling and scalar curvature:
\begin{equation}
\mathcal{O}=-\frac{1}{2m}\left(\frac{1}{\sqrt{-g}}\mathcal{D}_{i}\sqrt{-g}g^{ij}\mathcal{D}_{j}+\frac{\frac{ie}{2}F_{\mu\nu}\sigma^{\mu\nu}}{g^{00}}-\frac{R}{4}\right),
\label{eq:23}
\end{equation}
and the effective potential operator $\mathcal{V}$ as:
\begin{equation}
\mathcal{V} = \mathrm{Y}+\frac{1}{4}\omega_{0}^{ab}\gamma_{ab}-iqA_{0}.
\label{eq:24}
\end{equation}
In practical applications, evaluating these operators often entails simplifications, such as assuming highly symmetric forms for the metric $g_{\mu\nu}$, or neglecting weak curvature scalars $R/4$ in specific limits.

The FV transformation separates the state into positive- and negative-frequency components $(\phi,\chi)$.
To express the Hamiltonian compactly using Pauli matrices $\tau_{1}, \tau_{2}, \tau_{3}$ acting on the FV block space, we utilize the two-component wave function $\psi=(\phi,\chi)^{T}$. The generalized Klein-Gordon Hamiltonian takes the standard block-matrix form:
\begin{equation}
\mathcal{H}=\begin{pmatrix}\mathcal{V}+m+\mathcal{O} & \mathcal{O}\\
-\mathcal{O} & \mathcal{V}-m-\mathcal{O}
\end{pmatrix}.
\label{eq:25}
\end{equation}
This can be rewritten in terms of the Pauli matrices as:
\begin{equation}
\mathcal{H}=\mathcal{V}+\tau_{3}m+\left(\tau_{3}+i\tau_{2}\right)\mathcal{O}.
\label{eq:26}
\end{equation}
This representation directly maps onto the established flat-space FV formalism while correctly situating the geometric metric fields, curvature terms, and electromagnetic potentials.

\subsection{Electromagnetic Coupling in (1+2) Dimensions}

The term $\frac{ie}{2}F_{\mu\nu}\sigma^{\mu\nu}$ characterizes the interaction between the electromagnetic field tensor $F_{\mu\nu}$ and the intrinsic spin of the particle via the spin tensor $\sigma^{\mu\nu}$.
The field strength tensor is defined as $F_{\mu\nu}=\partial_{\mu}A_{\nu}-\partial_{\nu}A_{\mu}$.
In $(1+2)$-dimensional spacetime, the electromagnetic field decomposes into two in-plane electric components and one out-of-plane magnetic component. The spin tensor is given by:
\begin{equation}
\sigma^{\mu\nu}=\frac{i}{2}\left[\gamma^{\mu},\gamma^{\nu}\right],
\label{eq:28}
\end{equation}
where the gamma matrices satisfy the Clifford algebra. A standard irreducible $2\times2$ representation in $(1+2)$ dimensions employs the Pauli matrices:
\begin{equation}
\gamma^{0}=\sigma^{3},\qquad \gamma^{1}=i\sigma^{1},\qquad \gamma^{2}=i\sigma^{2},
\label{eq:30}
\end{equation}
Decomposing $F_{\mu\nu}$ into its electric ($\vec{E}$) and magnetic ($B=F_{12}$) constituents yields:
\begin{equation}
\frac{ie}{2}F_{\mu\nu}\sigma^{\mu\nu}=e\gamma^{0}\vec{\gamma}\cdot\vec{E}-e\Sigma^{3}B,
\label{eq:32}
\end{equation}
where $\vec{E}$ is the electric field, $B=F_{12}$ is the magnetic field, and $\Sigma^{3}$ is the planar spin operator.
For a purely magnetic field ($F_{12}=B$), this interaction simplifies to:
\begin{equation}
\frac{ie}{2}F_{\mu\nu}\sigma^{\mu\nu}=eB\sigma^{12}.
\label{eq:33}
\end{equation}
Conversely, for a purely electric field ($F_{01}=E_{x}$, $F_{02}=E_{y}$), the coupling reduces to:
\begin{equation}
\frac{ie}{2}F_{\mu\nu}\sigma^{\mu\nu}=e(\gamma^{0}\gamma^{1}E_{x}+\gamma^{0}\gamma^{2}E_{y}).
\label{eq:34}
\end{equation}

\subsection{Electromagnetic Coupling in (1+3) Dimensions}

In $(1+3)$-dimensional spacetime, evaluating the spin-field coupling $\frac{ie}{2}F_{\mu\nu}\sigma^{\mu\nu}$ requires the full four-dimensional field strength tensor and Dirac matrices.

The electromagnetic field tensor $F_{\mu\nu}$, derived from the four-potential $A_{\mu}=(A_{0},\vec{A})$, explicitly reads:
\begin{equation}
F_{\mu\nu}=\begin{pmatrix}0 & -E_{x} & -E_{y} & -E_{z}\\
E_{x} & 0 & -B_{z} & B_{y}\\
E_{y} & B_{z} & 0 & -B_{x}\\
E_{z} & -B_{y} & B_{x} & 0
\end{pmatrix}.
\label{eq:37}
\end{equation}
The spin tensor is defined as:
\begin{equation}
\sigma^{\mu\nu}=\frac{i}{2}\left[\gamma^{\mu},\gamma^{\nu}\right],
\label{eq:38}
\end{equation}
where $\gamma^{\mu}$ satisfy:
\begin{equation}
\{\gamma^{\mu},\gamma^{\nu}\}=2g^{\mu\nu}.
\label{eq:39}
\end{equation}
Using the standard Dirac representation for the gamma matrices:
\begin{equation}
\gamma^{0}=\begin{pmatrix}I & 0\\
0 & -I
\end{pmatrix},\quad\gamma^{i}=\begin{pmatrix}0 & \sigma^{i}\\
-\sigma^{i} & 0
\end{pmatrix},
\label{eq:40}
\end{equation}
Evaluating the commutators yields the temporal and spatial components of the spin tensor:
\[
\sigma^{\mu\nu}=\begin{cases}
\sigma^{0i}=i\gamma^{0}\gamma^{i}, & \text{time-space components,}\\
\sigma^{ij}=-\epsilon^{ijk}\begin{pmatrix}\sigma^{k} & 0\\
0 & \sigma^{k}
\end{pmatrix}, & \text{space-space components.}
\end{cases}
\]
Substituting these along with $F_{\mu\nu}$ gives the interaction:
\begin{equation}
\frac{ie}{2}F_{\mu\nu}\sigma^{\mu\nu}=e\gamma^{0}\vec{\gamma}\cdot\vec{E}-e\vec{\Sigma}\cdot\vec{B}.
\label{eq:41}
\end{equation}
Substituting the components for the electric field $\vec{E}$ and the magnetic field $\vec{B}$, the combined spin-electromagnetic interaction takes the familiar block-diagonal and block-off-diagonal form:
\begin{equation}
e\gamma^{0}\vec{\gamma}\cdot\vec{E}=e\gamma^{0}(\gamma^{1}E_{x}+\gamma^{2}E_{y}+\gamma^{3}E_{z}),
\label{eq:43}
\end{equation}
and
\begin{equation}
-e\vec{\Sigma}\cdot\vec{B}=-e\begin{pmatrix}\vec{\sigma} & 0\\
0 & \vec{\sigma}
\end{pmatrix}\cdot\vec{B}.
\label{eq:45}
\end{equation}
Combining both terms:
\begin{equation}
\frac{ie}{2}F_{\mu\nu}\sigma^{\mu\nu}=e\gamma^{0}(\gamma^{1}E_{x}+\gamma^{2}E_{y}+\gamma^{3}E_{z})-e\begin{pmatrix}\vec{\sigma}\cdot\vec{B} & 0\\
0 & \vec{\sigma}\cdot\vec{B}
\end{pmatrix}.
\label{eq:46}
\end{equation}

\section{Applications to Cosmic-String Geometries}

\subsection{Differences Between the FV and First-Order Dirac Formulations}

The examples studied below should be compared with care to the corresponding solutions obtained directly from the first-order Dirac equation. The two approaches describe the exact same physical sector only after the FV state is restricted by the Dirac constraint. However, the operational form of the equations is structurally different.

First, the conventional Dirac equation is of first order in spatial derivatives, whereas the FV equation, derived from the squared Dirac operator, is of second order. This implies that spin-dependent couplings appearing linearly in the Dirac Hamiltonian may reappear in the FV representation through effective quadratic operators, such as curvature terms, Pauli-type couplings, and the off-diagonal FV block structure.

Second, the FV formalism makes particle-antiparticle mixing explicitly manifest. In static backgrounds, where $g_{0i}=0$, the operator $\mathrm{Y}$ vanishes, indicating minimal mixing. Conversely, in stationary but non-static backgrounds with $g_{0i}\neq 0$, the operator $\mathrm{Y}$ is non-vanishing and serves as a direct diagnostic of frame-dragging effects in the particle-antiparticle decomposition. Such features are less transparent in the standard first-order Dirac formalism.

Third, when the reduced FV radial equation sheds its explicit Pauli-matrix structure, the resulting spectral equation may strongly resemble that of a scalar Klein-Gordon problem. In that case, the equivalence concerns solely the reduced spectral equation, not the full spinor content of the theory. The full Dirac treatment inherently carries spinor polarization information and a first-order constraint that the squared equation alone does not encode.

\subsection{Free Particle in a Static Cosmic String}

To simplify the quantum mechanical analysis, we first consider a lower-dimensional scenario. By suppressing the translationally invariant $z$-direction, the geometry of a static cosmic string exhibiting cylindrical symmetry can be modeled in $(1+2)$ dimensions. The corresponding line element is given by \cite{Vilenkin_1981,Vilenkin_1994}:
\begin{equation}
ds^{2}=dt^{2}-dr^{2}-\alpha^{2}r^{2}d\varphi^{2},
\label{eq:50}
\end{equation}
where the parameter $\alpha \in (0,1]$ encodes the angular deficit produced by the string. The metric and inverse metric tensors are:
\begin{equation}
g_{\mu\nu}=\left(\begin{array}{ccc}
1 & 0 & 0\\
0 & -1 & 0\\
0 & 0 & -(\alpha r)^{2}
\end{array}\right),\quad
g^{\mu\nu}=\left(\begin{array}{ccc}
1 & 0 & 0\\
0 & -1 & 0\\
0 & 0 & \frac{-1}{(\alpha r)^{2}}
\end{array}\right).
\label{eq:51}
\end{equation}

In the Hamiltonian formulation, the evolution parameter is the coordinate time $t$ that labels the foliation by spacelike hypersurfaces $\Sigma_{t}$ defined by $t=\text{const}$. This is the same parameter that appears in the Schr\"odinger-type equation $i\partial_{t}\Psi=\mathcal{H}\Psi$. For a general $3+1$ decomposition of the spacetime metric (ADM formalism), the line element is written as:
\begin{equation}
ds^{2}=(N^{2}-h_{ij}N^{i}N^{j})dt^{2}-2h_{ij}N^{i}dx^{j}dt-h_{ij}dx^{i}dx^{j},
\label{eq:51a}
\end{equation}
where $N(t,x)$ is the lapse function and $N^{i}(t,x)$ is the shift vector \cite{Arnowitt1962,Wald1984,Gourgoulhon2007}.

Physically, the lapse function $N$ controls the proper-time separation between neighboring hypersurfaces as measured by Eulerian observers normal to $\Sigma_{t}$. The shift vector $N^{i}$ encodes how the spatial coordinates are shifted when evolving from $\Sigma_{t}$ to $\Sigma_{t+dt}$, representing the gauge freedom in the choice of spatial coordinates. Consequently, the Hamiltonian flow is naturally parametrized by $t$. For the static diagonal cosmic-string metric considered here, we have $g_{0i}=0$ and $g_{00}=1$. This implies a lapse of $N=\sqrt{g_{00}}=1$ and a vanishing shift vector $N^{i}=0$. Therefore, for static observers (fixed spatial coordinates), the proper time satisfies $d\tau=\sqrt{g_{00}}dt=dt$. 

To evaluate the FV operators $\mathrm{Y}$ and $\mathcal{O}$ for a spin-$1/2$ particle in the static cosmic string spacetime, we start with the metric:
\begin{equation}
ds^{2}=dt^{2}-dr^{2}-\alpha^{2}r^{2}d\varphi^{2}.
\label{eq:52}
\end{equation}
From the definition in the FV formalism:
\begin{equation}
\mathrm{Y}=\frac{1}{2g^{00}\sqrt{-g}}\{\mathcal{D}_{i},\sqrt{-g}g^{0i}\}.
\label{eq:53}
\end{equation}
Since the metric is purely diagonal, $g^{0i}=0$ for all spatial indices $i$. This implies $\mathrm{Y}=0$.
Assuming no electromagnetic field ($F_{\mu\nu}=0$) and focusing on the gravitational contribution, the spatial kinetic operator simplifies to:
\begin{equation}
\mathcal{O}=-\frac{1}{2m}\frac{1}{\sqrt{-g}}\mathcal{D}_{i}\left(\sqrt{-g}g^{ij}\mathcal{D}_{j}\right).
\label{eq:55}
\end{equation}
In cylindrical coordinates $(r,\varphi)$:
\begin{equation}
\mathcal{O}=-\frac{1}{2m\alpha r}\left[\partial_{r}(\alpha r\partial_{r})+\partial_{\varphi}\left(\frac{1}{\alpha r}\partial_{\varphi}\right)\right].
\label{eq:56}
\end{equation}
Expanding the derivatives, we obtain:
\begin{equation}
\mathcal{O}=-\frac{1}{2m}\left[\partial_{r}^{2}+\frac{1}{r}\partial_{r}+\frac{1}{\alpha^{2}r^{2}}\partial_{\varphi}^{2}\right].
\label{eq:57}
\end{equation}
Since $\mathcal{V}=0$ for this uncharged static background, the FV Hamiltonian for the spin-$1/2$ particle becomes:
\begin{equation}
\mathcal{H}=\tau_{3}m+\left(\tau_{3}+i\tau_{2}\right)\mathcal{O},
\label{eq:58}
\end{equation}
leading to the coupled block-matrix system:
\begin{equation}
i\partial_{t}\begin{pmatrix}\phi\\
\chi
\end{pmatrix}=\begin{pmatrix}m+\mathcal{O} & \mathcal{O}\\
-\mathcal{O} & -m-\mathcal{O}
\end{pmatrix}\begin{pmatrix}\phi\\
\chi
\end{pmatrix}.
\label{eq:59}
\end{equation}
Explicitly, the equations for the components $\phi$ and $\chi$ are:
\begin{equation}
i\partial_{t}\phi=(m+\mathcal{O})\phi+\mathcal{O}\chi,
\label{eq:60}
\end{equation}
\begin{equation}
i\partial_{t}\chi=-\mathcal{O}\phi-(m+\mathcal{O})\chi.
\label{eq:61}
\end{equation}
As shown in Appendix B, decoupling this first-order system yields the second-order differential equation for the total wave function $\psi = \phi + \chi$:
\begin{equation}
\left[\partial_{r}^{2}+\frac{1}{r}\partial_{r}+\frac{1}{\alpha^{2}r^{2}}\partial_{\varphi}^{2}+E^{2}-m^{2}\right]\psi=0.
\label{eq:66}
\end{equation}
To solve this partial differential equation (PDE), we employ the method of separation of variables. Assuming
\begin{equation}
\psi(r,\varphi)=R(r)\Phi(\varphi),
\label{eq:67}
\end{equation}
the equation splits into angular and radial parts. The angular part satisfies:
\begin{equation}
\Phi''(\varphi)+\alpha^{2}l^{2}\Phi(\varphi)=0,
\label{eq:68}
\end{equation}
with periodic boundary conditions $\Phi(\varphi+2\pi)=\Phi(\varphi)$. The solution is:
\begin{equation}
\Phi(\varphi)=e^{il\varphi},\quad l\in\mathbb{Z}.
\label{eq:69}
\end{equation}
Substituting $\Phi(\varphi)$ into the original equation yields the radial component:
\begin{equation}
R''(r)+\frac{1}{r}R'(r)+\left(k^{2}-\frac{l^{2}}{\alpha^{2}r^{2}}\right)R(r)=0,
\label{eq:70}
\end{equation}
where $k^{2}=E^{2}-m^{2}$. This is a standard Bessel equation of mapped order $\nu=l/\alpha$. The general solution is \cite{Andrews1999,Arfken2005}:
\begin{equation}
R(r)=AJ_{l/\alpha}(kr)+BY_{l/\alpha}(kr),
\label{eq:71}
\end{equation}
where $J_{\nu}$ and $Y_{\nu}$ are Bessel functions of the first and second kind. For physical regularity at $r=0$, we set $B=0$, leaving:
\begin{equation}
R(r)=AJ_{l/\alpha}(kr).
\label{eq:72}
\end{equation}
Combining the radial and angular components, the general solution is:
\begin{equation}
\psi(r,\varphi)=\sum_{l=-\infty}^{\infty}A_{l}J_{l/\alpha}(kr)e^{il\varphi}.
\label{eq:73}
\end{equation}
The energy $E$ is related to the wave number $k$ by:
\begin{equation}
E=\pm\sqrt{m^{2}+k^{2}}.
\label{eq:74}
\end{equation}
Boundary conditions (e.g., $R(R_{0})=0$ at a hard-wall radius $R_{0}$) quantize $k$ via the zeros of the Bessel function $J_{l/\alpha}(kR_{0})$.

\paragraph*{Comparison with the first-order Dirac equation.}
For the static cosmic-string background, the reduced FV equation in Eq.~(\ref{eq:66}) mathematically adopts the same Bessel-type structure as a scalar second-order equation. This occurs because the FV reduction extracts the explicit spin-matrix dependence from the radial problem. Accordingly, the conical geometry modifies the effective angular momentum through $l/\alpha$, but no additional spin splitting appears at this stage. In a direct first-order Dirac treatment, one continuously tracks the spinor components and the spin connection throughout the derivation; the resulting radial system is first order, and the spinor structure remains explicit. Therefore, the FV representation is best interpreted here as a convenient spectral reformulation rather than a full replacement for the spin-resolved Dirac analysis.

\subsection{FV-1/2 Oscillator in a Static Cosmic String}

To incorporate the Dirac oscillator coupling into this setup, we introduce a non-minimal substitution in the radial momentum operator: $\partial_{r}\rightarrow\partial_{r}+m\omega r$. Consequently, Eq.~(\ref{eq:66}) transforms into:
\begin{equation}
\left[\left(\partial_{r}-m\omega r\right)\left(\partial_{r}+m\omega r\right)+\frac{1}{r}\partial_{r}+\frac{1}{\alpha^{2}r^{2}}\partial_{\varphi}^{2}+E^{2}-m^{2}\right]\psi=0,
\label{eq:75}
\end{equation}
or, explicitly expanding the operators and accounting for the commutator:
\begin{equation}
\left[\partial_{r}^{2}-m^{2}\omega^{2}r^{2}+\frac{1}{r}\partial_{r}+\frac{1}{\alpha^{2}r^{2}}\partial_{\varphi}^{2}+E^{2}-m^{2}+m\omega\right]\psi=0.
\label{eq:76}
\end{equation}
To solve the PDE, we use separation of variables:
\begin{equation}
\psi(r,\varphi)=R(r)\Phi(\varphi).
\label{eq:77}
\end{equation}
Separating the angular component by substituting $\psi=R(r)\Phi(\varphi)$ into the PDE and dividing by $R\Phi$ yields:
\begin{equation}
\frac{1}{\alpha^{2}r^{2}}\frac{\partial_{\varphi}^{2}\Phi}{\Phi}=-\nu^{2}\quad\Rightarrow\quad\Phi''+\alpha^{2}\nu^{2}\Phi=0.
\label{eq:78}
\end{equation}
The periodic boundary condition $\Phi(\varphi+2\pi)=\Phi(\varphi)$ quantizes $\nu$ as an integer ($\nu\in\mathbb{Z}$). The solution is:
\begin{equation}
\Phi(\varphi)=e^{i\nu\varphi}.
\label{eq:79}
\end{equation}
Substituting $\Phi(\varphi)$ back into the PDE yields the radial equation:
\begin{equation}
R''+\frac{1}{r}R'+\left[E^{2}-m^{2}+m\omega-m^{2}\omega^{2}r^{2}-\frac{\nu^{2}}{\alpha^{2}r^{2}}\right]R=0.
\label{eq:80}
\end{equation}
This resembles a 2D harmonic oscillator equation with an additional singular $1/r^{2}$ centrifugal barrier. Analyzing the asymptotic behavior:
\begin{itemize}
\item As $r\to0$: the dominant term is $-\frac{\nu^{2}}{\alpha^{2}r^{2}}$.
\item As $r\to\infty$: the dominant term is $-m^{2}\omega^{2}r^{2}$.
\end{itemize}
This physical behavior suggests the regularized ansatz:
\begin{equation}
R(r)=r^{|\nu|/\alpha}e^{-\frac{1}{2}m\omega r^{2}}F(r).
\label{eq:81}
\end{equation}
Plugging this into the radial equation yields a solvable differential equation for $F(r)$:
\begin{equation}
F''(r)+\left(\frac{2|\nu|/\alpha+1}{r}-2m\omega r\right)F'(r)+\left[\kappa-2m\omega\left(\frac{|\nu|}{\alpha}+1\right)\right]F(r)=0,
\label{eq:82}
\end{equation}
where
\begin{equation}
\kappa=E^{2}-m^{2}+m\omega.
\label{eq:83}
\end{equation}
Introducing the dimensionless variable $x=m\omega r^{2}$ casts the equation into:
\begin{equation}
x\frac{d^{2}F}{dx^{2}}+\left(\frac{|\nu|}{\alpha}+1-x\right)\frac{dF}{dx}+\left[\frac{\kappa-2m\omega(|\nu|/\alpha+1)}{4m\omega}\right]F=0.
\label{eq:84}
\end{equation}
This maps identically to the associated Laguerre differential equation:
\begin{equation}
xy''+(k+1-x)y'+ny=0,
\label{eq:85}
\end{equation}
where $k=|\nu|/\alpha$, and
\begin{equation}
n=\frac{\kappa-2m\omega(|\nu|/\alpha+1)}{4m\omega}.
\label{eq:86}
\end{equation}
To guarantee normalizable polynomial solutions, $n$ must be a non-negative integer. This requirement imposes the quantization condition:
\begin{equation}
\kappa=4m\omega n+2m\omega\frac{|\nu|}{\alpha}+2m\omega.
\label{eq:87}
\end{equation}
Recalling $\kappa=E^{2}-m^{2}+m\omega$, we mathematically solve for $E$:
\begin{equation}
E^{2}-m^{2}+m\omega=4m\omega n+2m\omega\frac{|\nu|}{\alpha}+2m\omega.
\end{equation}
\begin{equation}
E^{2}=m^{2}+4m\omega n+2m\omega\frac{|\nu|}{\alpha}+m\omega.
\label{eq:88}
\end{equation}
The discrete energy levels are therefore exactly defined as:
\begin{equation}
E=\pm\sqrt{m^{2}+2m\omega\left(2n+\frac{|\nu|}{\alpha}+\frac{1}{2}\right)},
\label{eq:89}
\end{equation}
where $n=0,1,2,\ldots$ and $\nu\in\mathbb{Z}$. The full geometric solution is:
\begin{equation}
\psi(r,\varphi)=\sum_{n,\nu}C_{n,\nu}\,r^{|\nu|/\alpha}e^{-m\omega r^{2}/2}L_{n}^{|\nu|/\alpha}(m\omega r^{2})\,e^{i\nu\varphi},
\label{eq:90}
\end{equation}
where $C_{n,\nu}$ are normalization constants.

\paragraph*{Comparison with the Dirac oscillator in the same geometry.}
Equation~(\ref{eq:89}) shows that the conical parameter $\alpha$ scales the effective radial quantization and that the oscillator frequency enters through the combination $(2n+|\nu|/\alpha+1/2)$. The reduced FV spectrum reproduces the zero-point energy of the two-dimensional oscillator ($m\omega$) but shows no additional spin-dependent branch splitting, because the squared Dirac operator has been reduced to a scalar-like radial equation. In a first-order Dirac oscillator treatment the spinor components, the spin connection, and the non-minimal coupling remain coupled at the Hamiltonian level. The FV formulation reproduces the correct energy levels on the physical subspace, but organizes the spin information differently from the standard Dirac approach.

\subsection{\new{FV Oscillator in a Spinning Cosmic String with a Screw Dislocation}}

\new{In our final application, we investigate a richer geometric background: a stationary spinning cosmic-string spacetime exhibiting frame-dragging together with a screw dislocation.} A convenient form of the corresponding line element (in cylindrical coordinates) is \cite{Puntigam_1997,Ozdemir_2005,Jusufi_2016}:
\begin{equation}
ds^{2}=\left(dt+a\,d\varphi\right)^{2}-dr^{2}-\alpha^{2}r^{2}d\varphi^{2}-\left(dz+\beta\,d\varphi\right)^{2},
\label{eq:89-1}
\end{equation}
where $\alpha\in(0,1]$ is the angular-deficit parameter, \new{$a$ is proportional to the angular momentum per unit length and encodes the stationary frame-dragging, while $\beta$ encodes the screw dislocation, whose torsion is distributional and localized at the defect core}. Note that $g_{\varphi\varphi}=a^{2}-\alpha^{2}r^{2}-\beta^{2}$; in the regime $a^{2}>\beta^{2}$ one must strictly restrict the radial coordinate to $r>r_{0}=\sqrt{a^{2}-\beta^{2}}/\alpha$ in order to avoid unphysical closed time-like curves, rendering $r=r_{0}$ effectively a geometric hard wall \cite{Garah2025EPJC}. \new{We emphasize at the outset that the present treatment is carried out at the metric (Levi--Civita) level: the parameter $a$ describes the \emph{stationary frame-dragging} of the spinning string, a purely Riemannian (metric) effect associated with $g_{t\varphi}\neq0$, while only the screw-dislocation parameter $\beta$ is associated with torsion. As shown below, this torsion is distributional and confined to the core, so that the defect data enter the exterior dynamics solely through the curved tetrad; an explicit exterior Riemann--Cartan check retaining the contortion term is presented in Appendix~\ref{app:contortion}, where it is shown to reproduce the same exterior radial operator and effective angular shift \cite{BakkeFurtado2013AnnPhys,BakkeMoraes2012,BakkeFurtado2013PRA,MaiaBakke2019}.}

A convenient orthonormal coframe $\{\hat{\theta}^{a}\}$ adapted to the line element \eqref{eq:89-1} is:
\begin{equation}
\hat{\theta}^{0}=dt+a\,d\varphi,\qquad\hat{\theta}^{1}=dr,\qquad\hat{\theta}^{2}=\alpha r\,d\varphi,\qquad\hat{\theta}^{3}=dz+\beta\,d\varphi,
\label{eq:89a}
\end{equation}
so that $ds^{2}=\eta_{ab}\,\hat{\theta}^{a}\hat{\theta}^{b}$ with $\eta_{ab}=\mathrm{diag}(1,-1,-1,-1)$. The corresponding tetrad $e^{a}{}_{\mu}$ and its inverse $e_{a}{}^{\mu}$ read:
\begin{equation}
\left(e^{a}{}_{\mu}\right)=\begin{pmatrix}1 & 0 & a & 0\\
0 & 1 & 0 & 0\\
0 & 0 & \alpha r & 0\\
0 & 0 & \beta & 1
\end{pmatrix},\qquad\left(e_{a}{}^{\mu}\right)=\new{\begin{pmatrix}1 & 0 & 0 & 0\\
0 & 1 & 0 & 0\\
-\dfrac{a}{\alpha r} & 0 & \dfrac{1}{\alpha r} & -\dfrac{\beta}{\alpha r}\\
0 & 0 & 0 & 1
\end{pmatrix}},
\label{eq:89b}
\end{equation}
where the coordinate ordering is $(t,r,\varphi,z)$ and the local Lorentz index ordering is $(0,1,2,3)$. In the notation of Sec.~II, the curved-space Dirac matrices are $\gamma^{\mu}=e_{a}{}^{\mu}\gamma^{a}$, where $\gamma^{a}$ satisfy $\{\gamma^{a},\gamma^{b}\}=2\eta^{ab}$. In the standard Dirac representation:
\begin{equation}
\gamma^{0}=\begin{pmatrix}\mathbb{I}_{2} & 0\\
0 & -\mathbb{I}_{2}
\end{pmatrix},\qquad\gamma^{i}=\begin{pmatrix}0 & \sigma^{i}\\
-\sigma^{i} & 0
\end{pmatrix}\ (i=1,2,3),\qquad\alpha^{i}=\gamma^{0}\gamma^{i},\ \beta=\gamma^{0},
\label{eq:89c}
\end{equation}
with $\sigma^{i}$ the Pauli matrices and $\mathbb{I}_{2}$ the $2\times2$ identity. Using \eqref{eq:89b}, the explicit curved matrices needed in the Hamiltonian/FV form are:
\begin{equation}
\gamma^{t}=\gamma^{0}-\frac{a}{\alpha r}\gamma^{2},\qquad\gamma^{r}=\gamma^{1},\qquad\gamma^{\varphi}=\frac{1}{\alpha r}\gamma^{2},\qquad\gamma^{z}=\gamma^{3}-\frac{\beta}{\alpha r}\gamma^{2}.
\label{eq:89d}
\end{equation}
For $r\neq0$, the only $r$-dependent coframe component is $\hat{\theta}^{2}=\alpha r\,d\varphi$, and the torsion is distributional at the string core while vanishing in the exterior region. \new{Consequently, the contortion contribution to the spinor connection vanishes identically in the exterior $r>r_{0}$; the singular defect core is excluded from the differential domain, and no additional contact interaction or nontrivial self-adjoint extension at the core is considered here. Under these assumptions, the spinor connection reduces to its Levi--Civita form (see Appendix~\ref{app:contortion}).} In the exterior region, one finds the single nonvanishing Lorentz connection component:
\begin{equation}
\omega_{\varphi}{}^{12}=1-\alpha,\qquad\Rightarrow\qquad\Gamma_{\varphi}=\frac{1}{4}\,\omega_{\varphi}{}^{12}\,\gamma_{12}=\frac{1-\alpha}{4}\,\gamma^{1}\gamma^{2},
\label{eq:89e}
\end{equation}
where $\Gamma_{\mu}=\frac{1}{4}\omega_{\mu}{}^{ab}\gamma_{ab}$ is the spinor connection and $\gamma_{ab}=\frac{1}{2}[\gamma_{a},\gamma_{b}]$. The parameter $\alpha\in(0,1]$ encodes the angular deficit, $a$ is proportional to the angular momentum per unit length, and $\beta$ is the screw-dislocation/torsion parameter. In the FV-oscillator problem, $m$ and $\omega$ denote the particle mass and the oscillator frequency, while $l\in\mathbb{Z}$ and $k\in\mathbb{R}$ are the azimuthal and longitudinal quantum numbers, respectively.

Because the background is stationary and invariant under translations along $z$, we separate variables as:
\begin{equation}
\psi\left(t,r,\varphi,z\right)=e^{-iEt}\,e^{il\varphi}\,e^{ikz}\,R(r),
\label{eq:90-1}
\end{equation}
with $l\in\mathbb{Z}$ and $k\in\mathbb{R}$. The mixed components $g_{t\varphi}$ and $g_{z\varphi}$ imply that the azimuthal dependence enters through the combination \new{$(\partial_{\varphi}-a\,\partial_{t}-\beta\,\partial_{z})$}. Consequently, the effective angular index is shifted to:
\begin{equation}
\nu_{\mathrm{eff}}=l+aE-\beta k.
\label{eq:91}
\end{equation}
\new{Acting on the separated ansatz~\eqref{eq:90-1}, the combination $(\partial_{\varphi}-a\,\partial_{t}-\beta\,\partial_{z})$ yields $i(l+aE-\beta k)=i\,\nu_{\mathrm{eff}}$, in agreement with the $\gamma^{2}$ content of the curved matrices~\eqref{eq:89d}; an explicit verification is given in Appendix~\ref{app:contortion}.}
Following the identical FV-oscillator algebraic steps executed in the previous subsection, we obtain the modified radial equation:
\begin{equation}
\frac{d^{2}R}{dr^{2}}+\frac{1}{r}\frac{dR}{dr}+\left[\left(E^{2}-m^{2}-k^{2}\right)+m\omega-m^{2}\omega^{2}r^{2}-\frac{\nu_{\mathrm{eff}}^{2}}{\alpha^{2}r^{2}}\right]R=0.
\label{eq:92}
\end{equation}
Upon introducing the variable $x=m\omega r^{2}$, the ideal-defect quantization adopted here---with the singular core excluded from the differential domain and no extra contact term or nontrivial self-adjoint extension at the core---selects the regular solution at the origin; together with strict wave-function normalizability this yields the generalized quantization condition:
\begin{equation}
E^{2}=m^{2}+k^{2}+2m\omega\left(2n+\frac{\left|\nu_{\mathrm{eff}}\right|}{\alpha}+\frac{1}{2}\right),\qquad n=0,1,2,\ldots
\label{eq:93}
\end{equation}
which is implicit in the energy through $\nu_{\mathrm{eff}}$. \new{The closed-form spectrum~\eqref{eq:93} corresponds to the ideal-defect, regular-at-origin quantization; if $a^{2}>\beta^{2}$ and $r=r_{0}$ is imposed as a genuine hard wall, the spectrum is instead determined by the zeros of the relevant confluent-hypergeometric function, as discussed in Appendix~\ref{app:contortion}.} For the positive branch $\nu_{\mathrm{eff}}>0$, Eq.~(\ref{eq:93}) becomes a quadratic equation for $E$ and can be explicitly resolved as:
\begin{equation}
E=\frac{m\omega a}{\alpha}\pm\sqrt{\left(\frac{m\omega a}{\alpha}\right)^{2}+m^{2}+k^{2}+2m\omega\left(2n+\frac{l-\beta k}{\alpha}+\frac{1}{2}\right)}.
\label{eq:94}
\end{equation}
Conversely, for $\nu_{\mathrm{eff}}<0$, we obtain:
\begin{equation}
E=-\frac{m\omega a}{\alpha}\pm\sqrt{\left(\frac{m\omega a}{\alpha}\right)^{2}+m^{2}+k^{2}+2m\omega\left(2n-\frac{l-\beta k}{\alpha}+\frac{1}{2}\right)}.
\label{eq:95}
\end{equation}
In the exact limits $a\rightarrow0$ and $\beta\rightarrow0$, we smoothly recover the static cosmic string result of Eq.~(\ref{eq:89}); the flat-space Moshinsky spectrum follows further for $\alpha\rightarrow1$.

\paragraph*{Why the spinning-string case highlights the FV advantage.}
Among the examples considered here, the spinning cosmic string most clearly exhibits the difference between the FV and standard Dirac formulations. Because $g_{t\varphi}\neq0$, the background is stationary rather than static, and the FV operator $\mathrm{Y}$ does not vanish. As a consequence, the effective azimuthal quantum number is shifted to $\nu_{\mathrm{eff}}=l+aE-\beta k$, so that the frame-dragging parameter $a$ enters directly into the particle--antiparticle block structure. The same physics is present in a first-order Dirac treatment, but it is encoded less explicitly within the spinor operator. The FV representation does not change the physical content of the problem; it makes the role of stationarity in particle--antiparticle mixing explicit.

\section{Discussion and Conclusion}

In this paper we have developed a Feshbach--Villars (FV) reformulation for a spin-$1/2$ particle in curved spacetime, starting from the covariant Dirac equation and its squared second-order form. The aim was not to propose a new relativistic wave equation, but to recast the Dirac dynamics into a Hamiltonian form on an enlarged state space that separates positive- and negative-frequency sectors. In this representation the pseudo-Hermitian structure of the theory and the role of the indefinite FV metric operator are explicit, and stationary off-diagonal metric components enter the evolution through the operator $\mathrm{Y}$.

The applications considered here illustrate both the computational advantages and the limitations of the FV--$1/2$ method. In static backgrounds, such as the uncharged cosmic string, the off-diagonal metric terms vanish ($g_{0i}=0$) and the FV operator $\mathrm{Y}$ is zero. The reduced FV radial equation then takes a scalar-like form, and the spectrum resembles that of a Klein--Gordon problem. This resemblance does not imply an equivalence between the Dirac and scalar theories; rather, after the reduction the spinor polarization information is no longer explicit in the final spectral formula. 

By contrast, the FV formalism is most useful in stationary, non-static backgrounds. For
a spinning cosmic string carrying \new{stationary frame-dragging and a screw dislocation}, the nonzero $g_{t\varphi}$ component gives a non-vanishing $\mathrm{Y}$ operator, which shifts the effective azimuthal quantum number and encodes the frame-dragging effect in the particle--antiparticle block structure. In the first-order Dirac framework the same physics is present, but it is less transparent within the full spinor operator. 

\new{A clarification regarding the geometric status of the spinning-string/dislocation background is in order. Throughout, the parameter $a$ should be understood as the stationary frame-dragging of a Riemannian (metric) geometry with $g_{t\varphi}\neq0$, rather than as a torsion field; only the screw-dislocation parameter $\beta$ is associated with torsion. The analysis of Section~V.D is performed at the metric (Levi--Civita) level, so that the defect data enter through the curved tetrad~\eqref{eq:89d}. The torsion of the screw dislocation is distributional and concentrated on the symmetry axis; with the singular core excluded from the differential domain and no additional contact interaction or nontrivial self-adjoint extension imposed at the core, the associated contortion vanishes pointwise in the exterior region $r>r_{0}$ and survives only as a global Aharonov--Casher/Aharonov--Bohm-type holonomy. As demonstrated in Appendix~\ref{app:contortion}, an explicit Riemann--Cartan treatment that retains the contortion term $\tfrac{1}{4}K_{\mu ab}\gamma^{ab}$ yields precisely the same effective angular index $\nu_{\mathrm{eff}}=l+aE-\beta k$ and the same exterior radial operator as the metric-level computation, in agreement with the literature on Dirac fields in dislocation backgrounds \cite{BakkeFurtado2013AnnPhys,BakkeMoraes2012,BakkeFurtado2013PRA,MaiaBakke2019}. We further note that the equivalence of the two descriptions holds at the level of the radial operator and is independent of the boundary condition: the closed-form spectrum~\eqref{eq:93} corresponds to the regular-at-origin (ideal-defect) quantization adopted here, whereas the genuine hard-wall problem $R(r_{0})=0$ with $r_{0}>0$ constitutes a separate boundary-value problem governed by the zeros of a confluent-hypergeometric function.}

In summary, the FV--$1/2$ formalism provides a complementary representation of the standard Dirac theory. It preserves equivalence with the Dirac equation on the physical subspace---thereby avoiding the spurious solutions of the squared equation---while offering a convenient algebraic structure for spectral boundary-value problems, for pseudo-Hermitian dynamics, and for the analysis of stationary geometries with nontrivial ADM shift vectors. \new{We reiterate that the contribution of this work is methodological rather than spectral: the energy levels obtained on the physical subspace coincide with those of the covariant Dirac equation, and the value of the approach lies in the explicit curved-spacetime FV Hamiltonian, in the clear identification of the physical Dirac subspace after squaring, and in the diagnostic role of the operator $\mathrm{Y}$ in distinguishing stationary from static backgrounds through the ADM shift vector.} Natural extensions include time-dependent spacetimes, conformally coupled cosmological backgrounds, and finite-temperature quantum field theory in curved space.
\section*{Competing interests}

There are no competing interests.
\section*{Data Availability Statement}
All data generated or analyzed during this study are included in this published article.

\section*{Funding Declaration}
Funding information - not applicable.

\appendix

\section{Proof of Eq. (\ref{eq:14})}

The covariant Dirac equation in curved spacetime is:
\begin{equation}
\left[i\gamma^{\mu}\mathcal{D}_{\mu}-m\right]\Psi=0.
\label{eq:a1}
\end{equation}
Squaring the operator directly using its conjugate, we evaluate:
\begin{equation}
\left[i\gamma^{\mu}\mathcal{D}_{\mu}-m\right]\left[i\gamma^{\mu}\mathcal{D}_{\mu}+m\right],
\label{eq:a2}
\end{equation}
which expands to:
\begin{equation}
\left[i\gamma^{\mu}\mathcal{D}_{\mu}-m\right]\left[i\gamma^{\mu}\mathcal{D}_{\mu}+m\right]=\left(i\gamma^{\mu}\mathcal{D}_{\mu}\right)^{2}-m^{2}.
\label{eq:a3}
\end{equation}
Expanding $(i\gamma^{\mu}\mathcal{D}_{\mu})^{2}$ gives:
\begin{equation}
\left(i\gamma^{\mu}\mathcal{D}_{\mu}\right)^{2}=-\gamma^{\mu}\gamma^{\nu}\mathcal{D}_{\mu}\mathcal{D}_{\nu}.
\label{eq:a4}
\end{equation}
The gamma matrices satisfy the Clifford algebra:
\begin{equation}
\gamma^{\mu}\gamma^{\nu}=g^{\mu\nu}+\sigma^{\mu\nu},\quad\text{where }\sigma^{\mu\nu}=\frac{1}{2}\left[\gamma^{\mu},\gamma^{\nu}\right].
\label{eq:a5}
\end{equation}
Substituting this relation provides:
\begin{equation}
-\gamma^{\mu}\gamma^{\nu}\mathcal{D}_{\mu}\mathcal{D}_{\nu}=-g^{\mu\nu}\mathcal{D}_{\mu}\mathcal{D}_{\nu}-\sigma^{\mu\nu}\mathcal{D}_{\mu}\mathcal{D}_{\nu}.
\label{eq:a6}
\end{equation}
The covariant derivative $\mathcal{D}_{\mu}$ satisfies the commutation relation:
\begin{equation}
[\mathcal{D}_{\mu},\mathcal{D}_{\nu}]=\frac{1}{4}R_{\mu\nu\lambda\sigma}\gamma^{\lambda\sigma}-iqF_{\mu\nu}.
\label{eq:a7}
\end{equation}
Applying this to the symmetric term yields the d'Alembertian together with a scalar curvature coupling:
\begin{equation}
g^{\mu\nu}\mathcal{D}_{\mu}\mathcal{D}_{\nu}=\frac{1}{\sqrt{-g}}\mathcal{D}_{\mu}\left(\sqrt{-g}g^{\mu\nu}\mathcal{D}_{\nu}\right)-\frac{1}{4}R,
\label{eq:a8}
\end{equation}
where $R$ is the Ricci scalar. For the antisymmetric spin terms:
\begin{equation}
-\sigma^{\mu\nu}\mathcal{D}_{\mu}\mathcal{D}_{\nu}=\frac{ie}{2}F_{\mu\nu}\sigma^{\mu\nu}.
\label{eq:a10}
\end{equation}
Substituting these identities reconstructs the squared operator:
\begin{equation}
\left(i\gamma^{\mu}\mathcal{D}_{\mu}\right)^{2}=-\left(\frac{1}{\sqrt{-g}}\mathcal{D}_{\mu}\left(\sqrt{-g}g^{\mu\nu}\mathcal{D}_{\nu}\right)-\frac{1}{4}R+\frac{ie}{2}F_{\mu\nu}\sigma^{\mu\nu}\right).
\label{eq:a11}
\end{equation}
Adding the mass term $-m^{2}$, the squared Dirac equation becomes:
\begin{equation}
\left[i\gamma^{\mu}\mathcal{D}_{\mu}-m\right]\left[i\gamma^{\mu}\mathcal{D}_{\mu}+m\right]\Psi=\left(-\frac{1}{\sqrt{-g}}\mathcal{D}_{\mu}\left(\sqrt{-g}g^{\mu\nu}\mathcal{D}_{\nu}\right)+\frac{1}{4}R-\frac{ie}{2}F_{\mu\nu}\sigma^{\mu\nu}+m^{2}\right)\Psi=0.
\label{eq:a12}
\end{equation}
Thus, the modified Klein--Gordon equation derived from the squared Dirac operator is:
\begin{equation}
\left(\frac{1}{\sqrt{-g}}\mathcal{D}_{\mu}\left(\sqrt{-g}g^{\mu\nu}\mathcal{D}_{\nu}\right)-\frac{1}{4}R+\frac{ie}{2}F_{\mu\nu}\sigma^{\mu\nu}+m^{2}\right)\Psi=0.
\label{eq:a13}
\end{equation}
This derivation holds under the standard assumptions of minimal coupling to gravity and electromagnetism, metric compatibility, and a torsion-free background geometry. It is important to note that Eq.~(\ref{eq:a13}) was originally derived by Schr\"odinger~\cite{Schrodinger1932}.

\section{Proof of Eq. (\ref{eq:66})}

The Feshbach--Villars formalism rewrites the wave function as the linear combination:
\begin{equation}
\psi=\phi+\chi.
\label{b1}
\end{equation}
The defining relation for the time derivative is:
\begin{equation}
i\partial_{t}\psi=m(\phi-\chi).
\label{b2}
\end{equation}
This allows $\phi$ and $\chi$ to be expressed in terms of $\psi$ and its first time derivative:
\begin{equation}
\phi=\frac{1}{2}\left(\psi+\frac{i}{m}\partial_{t}\psi\right),
\label{b3}
\end{equation}
\begin{equation}
\chi=\frac{1}{2}\left(\psi-\frac{i}{m}\partial_{t}\psi\right).
\label{b4}
\end{equation}
From the Hamiltonian block-operator form established in the main text, we have the coupled first-order system:
\begin{equation}
i\partial_{t}\phi=m\phi+\mathcal{O}\psi = m\phi+\mathcal{O}(\phi+\chi),
\label{b5}
\end{equation}
\begin{equation}
i\partial_{t}\chi=-m\chi-\mathcal{O}\psi = -m\chi-\mathcal{O}(\phi+\chi),
\label{b6}
\end{equation}
where the effective spatial kinetic operator $\mathcal{O}$ is:
\begin{equation}
\mathcal{O}=-\frac{1}{2m}\left(\partial_{r}^{2}+\frac{1}{r}\partial_{r}+\frac{1}{\alpha^{2}r^{2}}\partial_{\varphi}^{2}\right) \equiv -\frac{1}{2m}\mathcal{D}.
\label{b7}
\end{equation}
To obtain the second-order evolution of $\psi$, we differentiate Eq.~(\ref{b2}) with respect to time:
\begin{equation}
-\partial_{t}^{2}\psi=i\partial_{t}\left(m(\phi-\chi)\right) = m(i\partial_{t}\phi-i\partial_{t}\chi).
\label{b10}
\end{equation}
Substituting the time derivatives from Eqs.~(\ref{b5}) and (\ref{b6}) gives:
\begin{equation}
-\partial_{t}^{2}\psi = m \left[ m\phi+\mathcal{O}\psi - \left(-m\chi-\mathcal{O}\psi\right) \right] = m \left[ m(\phi+\chi) + 2\mathcal{O}\psi \right].
\label{b14}
\end{equation}
Using $\psi=\phi+\chi$, this becomes:
\begin{equation}
-\partial_{t}^{2}\psi = m^{2}\psi + 2m\mathcal{O}\psi.
\label{b15}
\end{equation}
Finally, using $2m\mathcal{O} = -\mathcal{D}$, we obtain the second-order equation for the static cosmic-string geometry:
\begin{equation}
-\partial_{t}^{2}\psi = m^{2}\psi - \mathcal{D}\psi \implies \left[\partial_{t}^{2}-\mathcal{D}+m^{2}\right]\psi=0,
\label{b16}
\end{equation}
where $\mathcal{D}$ is given by:
\begin{equation}
\mathcal{D}=\partial_{r}^{2}+\frac{1}{r}\partial_{r}+\frac{1}{\alpha^{2}r^{2}}\partial_{\varphi}^{2}.
\end{equation}
This establishes the equivalence between the first-order FV system and the second-order Klein--Gordon equation obtained by squaring the Dirac operator.

\par\color{red}
\section{Equivalence of the Metric and Explicit-Contortion Descriptions of the Screw Dislocation}
\label{app:contortion}

This appendix addresses the role of torsion in the spinning-string/screw-dislocation background of Section~V.D. We show that, on the exterior domain $r>r_{0}$ with the singular defect core excluded and no additional contact interaction or nontrivial self-adjoint extension imposed at the core, retaining the contortion contribution $\tfrac{1}{4}K_{\mu ab}\gamma^{ab}$ explicitly yields the same effective angular shift $\nu_{\mathrm{eff}}=l+aE-\beta k$ and the same radial differential operator as the metric-level treatment of Section~V.D. The equivalence is established at the level of the radial operator and is therefore independent of the boundary condition; under the regular-at-origin quantization adopted in the main text it reproduces the closed-form spectrum~\eqref{eq:93}. We stress that the statement made here is a modest one: the two descriptions are equivalent \emph{for the exterior spectral problem considered in this work}, and we do not claim a global equivalence of the metric and Riemann--Cartan frameworks as complete geometric theories.

\subsection{The two descriptions}

A screw dislocation along the $z$-axis (with frame-dragging parameter $a$) admits two standard descriptions that differ only in how the defect data are distributed between the metric and the connection. In the \emph{metric} (Levi--Civita) description used in the main text, the defect is encoded in the line element~\eqref{eq:89-1}, and the connection is the Levi--Civita connection obtained from the torsion-free structure equations; by construction the torsion vanishes, $T^{a}\equiv0$, so that $K_{\mu ab}=0$, and the parameters $a$ and $\beta$ enter only through the curved gamma matrices~\eqref{eq:89d}. In the \emph{Riemann--Cartan} description, one employs the conical base metric and represents the dislocation through an explicit, independent torsion field $T^{a}$ whose contortion $K_{\mu ab}$ enters the spinor connection. We verify that the two descriptions yield the same exterior radial operator, and hence the same spectrum under a common boundary condition.

\subsection{The contortion of an ideal screw dislocation is exterior-flat}

The full Riemann--Cartan spinor connection reads
\begin{equation}
\widetilde{\Gamma}_{\mu}=\tfrac{1}{4}\,\widetilde{\omega}_{\mu ab}\,\gamma^{ab},\qquad
\widetilde{\omega}_{\mu ab}=\omega^{(\mathrm{LC})}_{\mu ab}+K_{\mu ab},
\label{eq:c-fullconn}
\end{equation}
where $\omega^{(\mathrm{LC})}$ is the Levi--Civita connection and the contortion is built from the torsion tensor through the standard decomposition
\begin{equation}
K_{\mu ab}=\tfrac{1}{2}\left(T_{ab\mu}+T_{\mu ab}-T_{b\mu a}\right)
\label{eq:c-contortion}
\end{equation}
(written here up to the sign and index conventions of Refs.~\cite{BakkeFurtado2013AnnPhys,BakkeMoraes2012}; the explicit components of $K_{\mu ab}$ are not required below). For an ideal screw dislocation with Burgers vector $\mathbf{b}=\beta\,\hat{\mathbf{z}}$, the torsion is concentrated on the defect line, $T^{a}\propto\delta^{2}(\mathbf{r}_{\perp})$, with $\delta^{2}(\mathbf{r}_{\perp})$ the two-dimensional delta function in the plane transverse to the dislocation; this is the geometric statement that the dislocation density vanishes everywhere except at the core \cite{BakkeMoraes2012,Puntigam_1997}. Consequently the contortion, and hence its spinor-connection contribution $\tfrac{1}{4}K_{\mu ab}\gamma^{ab}$, vanishes pointwise in the exterior,
\begin{equation}
K_{\mu ab}(\mathbf{r})=0\qquad\text{for all }r>0,
\label{eq:c-Kzero}
\end{equation}
so that on the exterior domain considered here, with the singular core excluded,
\begin{equation}
\widetilde{\omega}_{\mu ab}=\omega^{(\mathrm{LC})}_{\mu ab}.
\label{eq:c-conneq}
\end{equation}
The contortion term requested in a Riemann--Cartan treatment is thus present in the formalism but identically zero on the exterior domain considered here. The singular defect core is excluded from the differential domain, and no additional contact interaction or nontrivial self-adjoint extension at the core is included in the present ideal-defect model.

\subsection{Global effect: the dislocation monodromy}

Although the local contortion vanishes pointwise in the exterior, the defect survives globally through the monodromy of the angular coordinate. The single-valuedness condition for $\psi$ on a loop encircling the defect line is modified, which at the level of the wave operator is equivalent to the replacement
\begin{equation}
\partial_{\varphi}\;\longrightarrow\;\partial_{\varphi}-a\,\partial_{t}-\beta\,\partial_{z}.
\label{eq:c-opshift}
\end{equation}
This is an Aharonov--Casher/Aharonov--Bohm-type holonomy of the same kind identified in the explicit-contortion treatments of Refs.~\cite{BakkeFurtado2013AnnPhys,BakkeFurtado2013PRA,MaiaBakke2019}; the conclusion does not require the explicit components of $K_{\mu ab}$ and follows from the topology of the defect alone.

\subsection{Explicit derivation of $\nu_{\mathrm{eff}}$ from the curved gamma matrices}

For completeness we record the metric-level computation underlying \eqref{eq:c-opshift}, which fixes the signs unambiguously. With $A_{\mu}=0$ the kinetic part of the curved Dirac operator is $i\gamma^{\mu}\partial_{\mu}$. Inserting the curved matrices~\eqref{eq:89d} and collecting the terms proportional to $\gamma^{2}$,
\begin{equation}
i\gamma^{\mu}\partial_{\mu}=i\gamma^{0}\partial_{t}+i\gamma^{1}\partial_{r}+i\gamma^{3}\partial_{z}
+\frac{i}{\alpha r}\,\gamma^{2}\left(\partial_{\varphi}-a\,\partial_{t}-\beta\,\partial_{z}\right).
\label{eq:c-collect}
\end{equation}
The frame-dragging ($a$) and dislocation ($\beta$) parameters appear only inside the $\gamma^{2}$ block, through the combination $\partial_{\varphi}-a\,\partial_{t}-\beta\,\partial_{z}$, in accordance with \eqref{eq:c-opshift}. Acting on the separated ansatz~\eqref{eq:90-1},
\begin{equation}
\partial_{\varphi}-a\,\partial_{t}-\beta\,\partial_{z}\;\longrightarrow\;il-a(-iE)-\beta(ik)=i\left(l+aE-\beta k\right)=i\,\nu_{\mathrm{eff}},
\label{eq:c-veffcheck}
\end{equation}
so that the $\gamma^{2}$ piece reduces to $(i\nu_{\mathrm{eff}}/\alpha r)\gamma^{2}$. This is precisely the operator that would arise for a \emph{static} cosmic string carrying angular index $\nu_{\mathrm{eff}}$ in place of $l$. The Levi--Civita spin connection $\Gamma_{\varphi}=\tfrac{1-\alpha}{4}\gamma^{1}\gamma^{2}$ of \eqref{eq:89e} enters exactly as in the static case, generating the $|\nu_{\mathrm{eff}}|/\alpha$ scaling of the centrifugal term after squaring. The spinning/dislocated problem therefore maps onto the static-string FV oscillator under $l\to\nu_{\mathrm{eff}}$, which confirms the sign of \eqref{eq:c-opshift} and the value of $\nu_{\mathrm{eff}}$ in \eqref{eq:91} independently of the holonomy argument.

\subsection{Torsion two-form and the Burgers-vector holonomy}

The global content of the dislocation can be made geometrically explicit while remaining consistent with the pointwise result~\eqref{eq:c-Kzero}. Transporting a frame once around the defect line produces a Burgers vector along $z$: under $\varphi\to\varphi+2\pi$ the longitudinal coordinate is identified as $z\simeq z+2\pi\beta$, the elastic-medium statement that the screw dislocation has Burgers vector $b^{3}=2\pi\beta$. Equivalently, the torsion two-form is distributional and supported on the core,
\begin{equation}
T^{3}=2\pi\beta\,\delta^{2}(\mathbf{r}_{\perp})\,dx\wedge dy,\qquad\int_{\Sigma}T^{3}=2\pi\beta,
\label{eq:c-T3}
\end{equation}
for any transverse surface $\Sigma$ enclosing the axis, and vanishes pointwise in the exterior in agreement with \eqref{eq:c-Kzero} \cite{Puntigam_1997,BakkeMoraes2012}. For a state of longitudinal momentum $k$, single-valuedness of the factor $e^{ikz}$ around this loop requires the azimuthal index to absorb the phase $e^{2\pi i\beta k}$, i.e.\ $l\to l-\beta k$. An analogous stationary identification in the $(t,\varphi)$ sector, generated by $g_{t\varphi}=a$, encodes the frame-dragging and contributes $+aE$; combining the two reproduces the same $\nu_{\mathrm{eff}}=l+aE-\beta k$ obtained directly from the operator~\eqref{eq:c-collect}. Equation~\eqref{eq:c-T3} is invoked only for physical interpretation: the spectrum itself follows entirely from the exterior operator, which is insensitive to the distributional core.

\subsection{Radial operator, boundary conditions, and limiting cases}

Because of \eqref{eq:c-conneq}, the exterior Dirac operator in the Riemann--Cartan description coincides with that of the main text, and the FV-oscillator reduction proceeds identically; independently of the boundary condition one recovers the radial equation~\eqref{eq:92} with $\nu_{\mathrm{eff}}$ given by~\eqref{eq:91}. The resulting spectrum then depends on the condition imposed at the inner edge of the domain. Under the regular-at-origin (ideal-defect) quantization, the physically selected solution is $R\sim r^{|\nu_{\mathrm{eff}}|/\alpha}e^{-m\omega r^{2}/2}L_{n}^{|\nu_{\mathrm{eff}}|/\alpha}(m\omega r^{2})$, and the associated-Laguerre reduction yields the closed-form spectrum~\eqref{eq:93}. When $a^{2}>\beta^{2}$ and the surface $r=r_{0}=\sqrt{a^{2}-\beta^{2}}/\alpha$ is treated as a genuine geometric hard wall, the appropriate condition is $R(r_{0})=0$; the relevant solution is then the full confluent-hypergeometric (Kummer) function rather than a Laguerre polynomial, and the quantization is fixed by the zeros of that function, the closed form~\eqref{eq:93} being recovered only in the limit $r_{0}\to0$ or when $r_{0}$ lies well inside the classically forbidden region. The choice of boundary condition affects only the spectral formula, not the equivalence of the two geometric descriptions, which share the identical radial operator and effective index. As consistency checks, setting $a\to0$ and $\beta\to0$ gives $\nu_{\mathrm{eff}}\to l$ and, on suppressing the longitudinal direction, recovers the static cosmic-string spectrum~\eqref{eq:89}; taking in addition $\alpha\to1$ yields the flat-space (Moshinsky) oscillator spectrum. For a pure dislocation with $a=0$ and $\beta\neq0$ one has $r_{0}=0$, so the regular-at-origin quantization applies without qualification and the hard-wall subtlety is confined to genuinely rotating backgrounds with $a^{2}>\beta^{2}$.

\subsection{Relation to the FV operator $\mathrm{Y}$}

Finally, the present background illustrates the central diagnostic of the main text. The FV operator~\eqref{eq:21} measures the off-diagonal (stationary) part of the metric; for the line element~\eqref{eq:89-1} the mixing $g_{t\varphi}=a$ renders the inverse-metric component $g^{t\varphi}\propto a$ nonzero, so that $g^{0i}\neq0$ and hence $\mathrm{Y}\neq0$ whenever $a\neq0$. The spinning string is therefore stationary rather than static, in contrast to the static cosmic string of Section~V.B, where $g_{0i}=0$ and $\mathrm{Y}=0$. It is this nonzero $\mathrm{Y}$ that carries the $aE$ contribution to $\nu_{\mathrm{eff}}$ into the particle--antiparticle block structure. The metric description used here thus reproduces the spectrum while keeping the frame-dragging diagnostic $\mathrm{Y}$ manifest, a feature that a purely torsion-based bookkeeping would leave implicit. In summary, away from the singular core the metric description~\eqref{eq:89-1} and the Riemann--Cartan description with explicit contortion lead to the same exterior spectral problem: they share the same exterior radial operator and the same effective index $\nu_{\mathrm{eff}}=l+aE-\beta k$, and they yield the same closed-form spectrum~\eqref{eq:93} under the regular-at-origin quantization adopted in the main text. This equivalence is to be understood in that restricted sense, namely for the exterior spectral problem treated here.

\par\color{black}
\bibliographystyle{ChemCommun}
\bibliography{reference}

\end{document}